\documentclass[12pt]{article}
\usepackage{epsfig,cite}

\def\ra{\rightarrow}
\newcommand{\ee}       {\mbox{$e^+ e^-$}}

\begin{document}
\bibliographystyle{hunsrt} 

\pagestyle{empty}

\renewcommand{\thefootnote}{\fnsymbol{footnote}}


\begin{flushright}
{\small
SLAC--PUB--10862\\
November 2004\\}
\end{flushright}

\vspace{.8cm}

\begin{center}
{\large\bf Experimental Aspects of Higgs Physics at the ILC\footnote{Work supported
by Department of Energy contract  DE--AC02--76SF00515.}}

\vspace{1cm}

Timothy L. Barklow\\
Stanford Linear Accelerator Center, Stanford University \\
2575 Sand Hill Road, Menlo Park, CA 94025 USA\\

\end{center}

\vfill

\begin{center}
{\large\bf
Abstract }
\end{center}

\begin{quote}
Recent progress in  Higgs boson studies for the
International $\ee$ Linear Collider (ILC) is reported. 
These studies
include extended simulations of the measurement of the Higgs mass, measurements of 
the Higgs boson branching ratios at higher center-of-mass energies, and methods for extracting the Higgs boson 
self-coupling.  
Also, the interplay between the LHC and the ILC in the
measurement of the top Yukawa coupling and in the extraction of the
supersymmetric Higgs sector parameters is discussed.
\end{quote}

\vfill

\begin{center}
\textit{Invited talk presented at International Conference On Linear Colliders (LCWS 04)} \\
\textit{Paris, France} \\
\textit{April 19 - April 24, 2004} 
\end{center}

\newpage

\pagestyle{plain}

\section{Introduction}

With an $\ee$  center of mass energy in the range of 90 to 1000~GeV, a well defined initial state, 
and a clean experimental environment, the International $\ee$ Linear Collider (ILC)
will be an ideal accelerator at which  
to study the properties of Higgs bosons.  
An extensive literature~\cite{Murayama:1996ec,Aguilar-Saavedra:2001rg,Abe:2001np,Abe:2001gc,Dawson:2004xz} 
exists detailing how the ILC can measure the masses, widths, couplings, and quantum
numbers of Higgs bosons in a model independent
manner with 
high precision.
  In this paper recent progress on experimental aspects of ILC Higgs boson physics
will be presented.  The discussion includes beam-related systematic errors in Higgs mass measurements,
hadronic Higgs decays, 
Higgs boson phenomenology at $\sqrt{s}=1000$~GeV, 
and LHC/ILC synergy.

\section{Beam Related Systematic Errors in the Higgs Mass Measurement}

     The Higgs mass will be measured in the process $\ee\ra Zh$ 
using the recoil mass technique
and the direct multi-jet technique if the fully hadronic decay branching ratio is high enough.
In the recoil mass technique the mass of the system opposite a $Z$ boson decaying to $\ee$ or $\mu^+\mu^-$
is measured without regard to the Higgs decay.  In the direct multi-jet technique, jets from the Higgs
decay are combined with the jets or leptons from the $Z$ boson decay in a kinematical fit, and the mass is extracted from 
the fitted four-vectors.

In both techniques
a kinematical constraint utilizing the beam energy is employed.  Systematic errors in the measurement of the beam energy scale
and the differential luminosity distribution
therefore contribute to the total Higgs mass error.

\subsection{Differential luminosity measurement}

The effects of beamstrahlung are described by a double differential luminosity distribution ${\rm d}^2L/{\rm d}x_1{\rm d}x_2$, where $x_1$ 
and $x_2$ are the energy fractions of the electrons and positrons.  
Most analyses utilize the acollinearity distribution of Bhabha events to reconstruct this distribution.
Studies indicate that under idealized
conditions where the function  ${\rm d}^2L/{\rm d}x_1{\rm d}x_2$ factorizes into two identical one-dimensional
distributions of the form $f(x_i) = a_0\delta(1-x_i)+a_1x_i^{a_2}(1-x_i)^{a_3}$, the differential luminosity can be measured to an accuracy of 1\%,
assuming 3 ${\rm fb}^{-1}$  at $\sqrt{s}=500$~GeV~\cite{Moenig:2000}.

          In a study of beam related systemtatic errors on Higgs mass measurements, Raspereza has shown that a 10\% measurement error of
the differential luminosity parameters $a_i$ leads to a 10 MeV error on the Higgs mass when the direct mult-jet technique is used~\cite{Raspereza:2004}.

\subsection{Beam energy scale}

The beam energy scale can be measured to an accuracy of about 200 ppm by  combining data from beam energy spectrometers  upstream and downstream of the interaction point 
with beam energy estimates from physics processes such as $\ee \ra \gamma Z,\ ZZ,\ \ee$ and $\mu^+\mu^-$~\cite{Woods:2004mf}.   

The dependence of the Higgs mass error on the beam energy scale error has been studied by several groups.  Raspereza studied the direct mult-jet
technique assuming a 120~GeV Standard Model Higgs boson and found
$\delta M_H/\delta {\rm E_{cm}}=0.5 $ for the $bbll$ final state and $\delta M_H/\delta {\rm E_{cm}}=0.4 $ for the
$bbqq$ final state.  Due to the fact that no Higgs decay information is utilized, the recoil mass technique has a much stronger 
depedence on the beam energy measurement with $\delta M_H/\delta {\rm E_{cm}}=2.9 $.
  
   The statistical 
error and energy scale systematic error for a 120~GeV standard model Higgs boson are summarized in Table~\ref{tab:syst}.

\begin{table}[htbp]
\center{\begin{tabular}{c c c}
technique   & statistical error (GeV)   & energy scale systematic error (GeV)   \\ \hline
recoil mass & 0.117 & 0.200  \\
$ZH \ra l^+l^-b\bar{b}$ & 0.072 & 0.035  \\
$ZH \ra q\bar{q} b\bar{b}$ & 0.046 & 0.028  \\ \hline
combined  & 0.037 & 0.027  \\ \hline
\end{tabular}
\caption{\label{tab:syst}\it 
Statistical and energy scale systematic errors for the Higgs mass measurement at the ILC 
assuming a standard model Higgs with $M_H=120$~GeV, $\sqrt{s}=350$~GeV, 500 fb$^{-1}$ luminosity, 
and a 200 ppm center-of-mass energy scale error.  
The combined energy scale systematic error includes the effects of correlations between the three measurements.
}}
\end{table}

\section{Hadronic Branching Ratio Measurement}

For some time European and American working groups 
have come to different
conclusions about how well hadronic Higgs branching ratios can  be measured~\cite{Aguilar-Saavedra:2001rg,Abe:2001np}.   
This is illustrated in the last two columns of Table~\ref{tab:comp},
where the branching ratio error estimates from the TESLA TDR are displayed alongside those reported at Snowmass 2001.
Recently, a new analysis
presented by Thorsten Kuhl at LCWS 2004~\cite{Kuhl:2004} 
splits the difference, as shown in the first column of Table~\ref{tab:comp} .

\begin{table}[htbp]
\center{\begin{tabular}{c|c c c}
           & LCWS       & TESLA TDR         & Snowmass     \\ 
selection  &      2004  &           (2001)  &          2001\\ \hline
$\Delta(\sigma\times BR(H\rightarrow b\overline{b}))/(\sigma\times BR(bb))_{SM}$ & 1.0$\%$  &0.9$\%$&1.6$\%$\\ 
$\Delta(\sigma\times BR(H\rightarrow c\overline{c}))/(\sigma\times BR(cc))_{SM}$ & 12.0$\%$ &8.0$\%$&19.0$\%$\\ 
$\Delta(\sigma\times BR(H\rightarrow gg))/(\sigma\times BR(gg))_{SM}$ & 8.2$\%$  &5.1$\%$&10.4$\%$ \\ 
\end{tabular}
\caption{\label{tab:comp}\it 
Comparison of the relative errors obtained with Kuhl's new  analysis and previous analyses from the TESLA TDR and the Snowmass 2001 workshop.}}
\end{table}

Compared to the 2001 TESLA TDR results, the new analysis by Kuhl uses an improved version of the TESLA detector fast Monte Carlo program SIMDET~\cite{Pohl:2002vk}.  
Results on tracking efficiency and impact parameter resolution from studies using the full TESLA detector Monte Carlo and emulated pattern recognition 
have been included in the new version of SIMDET.  These newly incorporated effects tend to worsen the charm tagging capabilities of the detector.
On the other hand the LCWS 2004 errors are better than those reported at Snowmass 2001 due to the superiority of 
maximum likelihood analyses of neural net variables over cuts-based analyses.

\section{Studies at {\boldmath $\sqrt{{\rm s}}=1000\ {\rm GeV}$}}
\subsection{Higgs self-coupling}

       The Higgs potential $V(\eta_H)$ is probed through measurements of the triple and quartic Higgs self-couplings $\lambda$ and $\tilde{\lambda}$ :
\begin{displaymath}
V(\eta_H)=\frac{1}{2}M_H^2\eta_H^2+\lambda v\eta_H^3+\frac{1}{4}\tilde{\lambda}\eta_H^4\ \ ,
\end{displaymath}
where $\eta_H$ is the Higgs boson field and $v$ is the Higgs vacuum expectation value. 
In the Standard Model $\lambda=\tilde{\lambda}=M_H^2/(2v^2)$, so that a comparison of the measured values of $\lambda$, $\tilde{\lambda}$ and $M_H$
will constitute an important test of electroweak symmetry breaking models.

       The triple Higgs self-coupling $\lambda$ will be measured at the ILC at $\sqrt{s}=500$~GeV in the Higgstrahlung process
$\ee\ra ZH^*\ra ZHH$.  
Assuming 500~fb$^{-1}$ luminosity, studies have shown that an accuracy of $\delta \lambda/\lambda=0.28$ can
be achieved 
for a Higgs mass of 120~GeV~\cite{Battaglia:2001nn}.

Recently, the possibility of measuring the triple Higgs coupling at the ILC at $\sqrt{s}=1000$~GeV using 
the WW fusion process $\ee\ra \nu_e \bar{\nu}_e W^*W^* \ra \nu_e \bar{\nu}_e H^* \ra \nu_e \bar{\nu}_e H H$
  has been considered~\cite{Yasui:2002se}.   
Assuming 1000~fb$^{-1}$ luminosity and 80\% left-handed electron polarization, 
a study of $\ee\ra \nu_e \bar{\nu}_e H H\ra \nu_e \bar{\nu}_e b\bar{b}b\bar{b}$
found that a triple Higgs coupling accuracy of 
$\delta \lambda/\lambda\approx 0.12$ could be achieved\cite{Yamashita:2004}.  Further improvement is expected by 
extending the analysis to decay topologies other than $HH\ra b\bar{b}b\bar{b}$.

\subsection{Higgs branching ratios}

CLIC studies have demonstrated that Higgs production through the $WW$ fusion process $\ee\ra\nu_e\bar{\nu}_e H$
at $\sqrt{s}=3000$~GeV can be used to probe rare Higgs decays~\cite{Battaglia:2001vf,Battaglia:2002gq}.  A recent study
has shown that such decays can also be probed through $WW$ fusion
at the ILC at $\sqrt{s}=1000$~GeV~\cite{Barklow:2003hz}.  

\begin{figure}[ht]
     \begin{center}
     \begin{tabular}{cc}
     \mbox{\epsfig{file=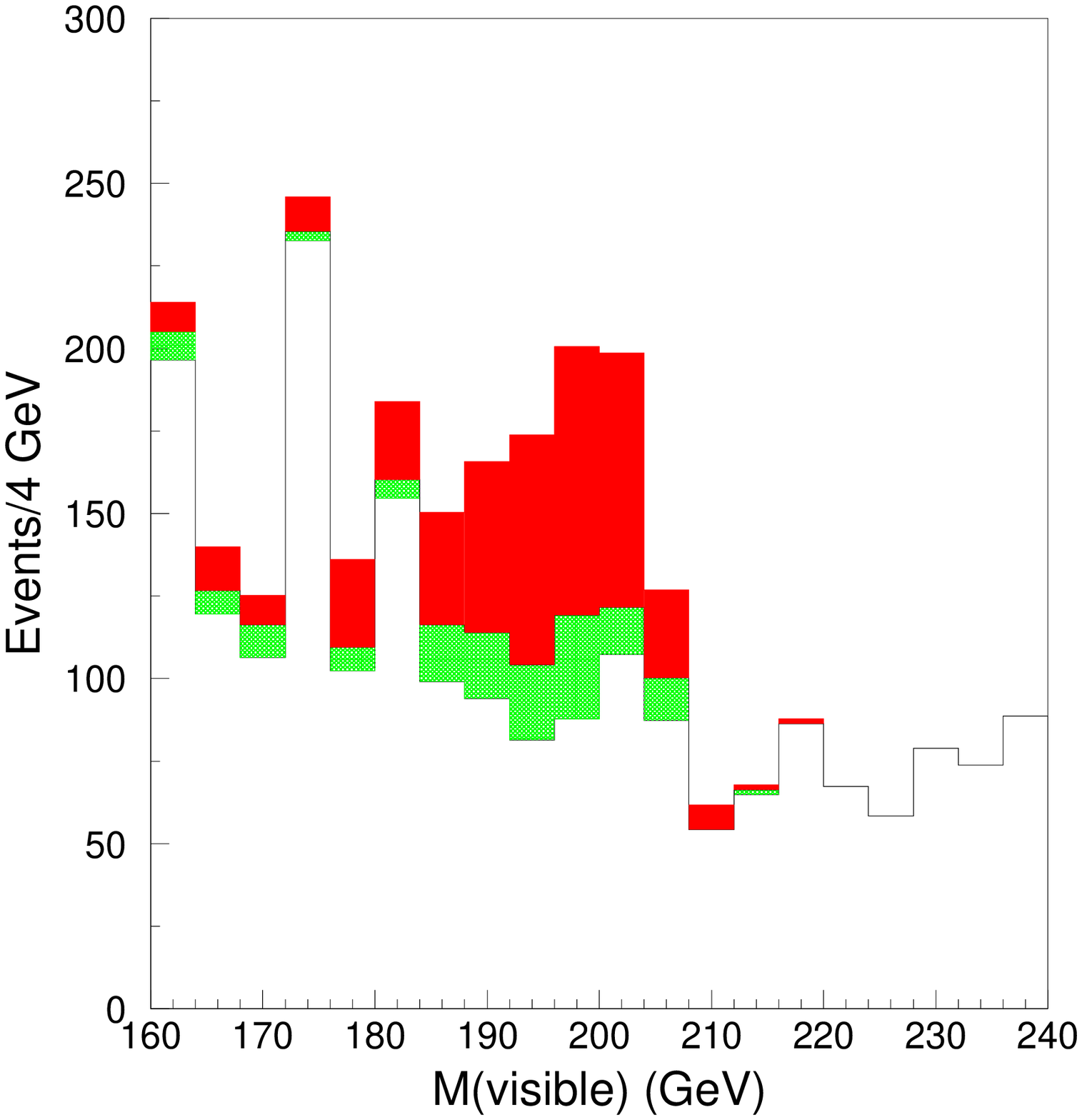,width=6.0cm}}&
     \mbox{\epsfig{file=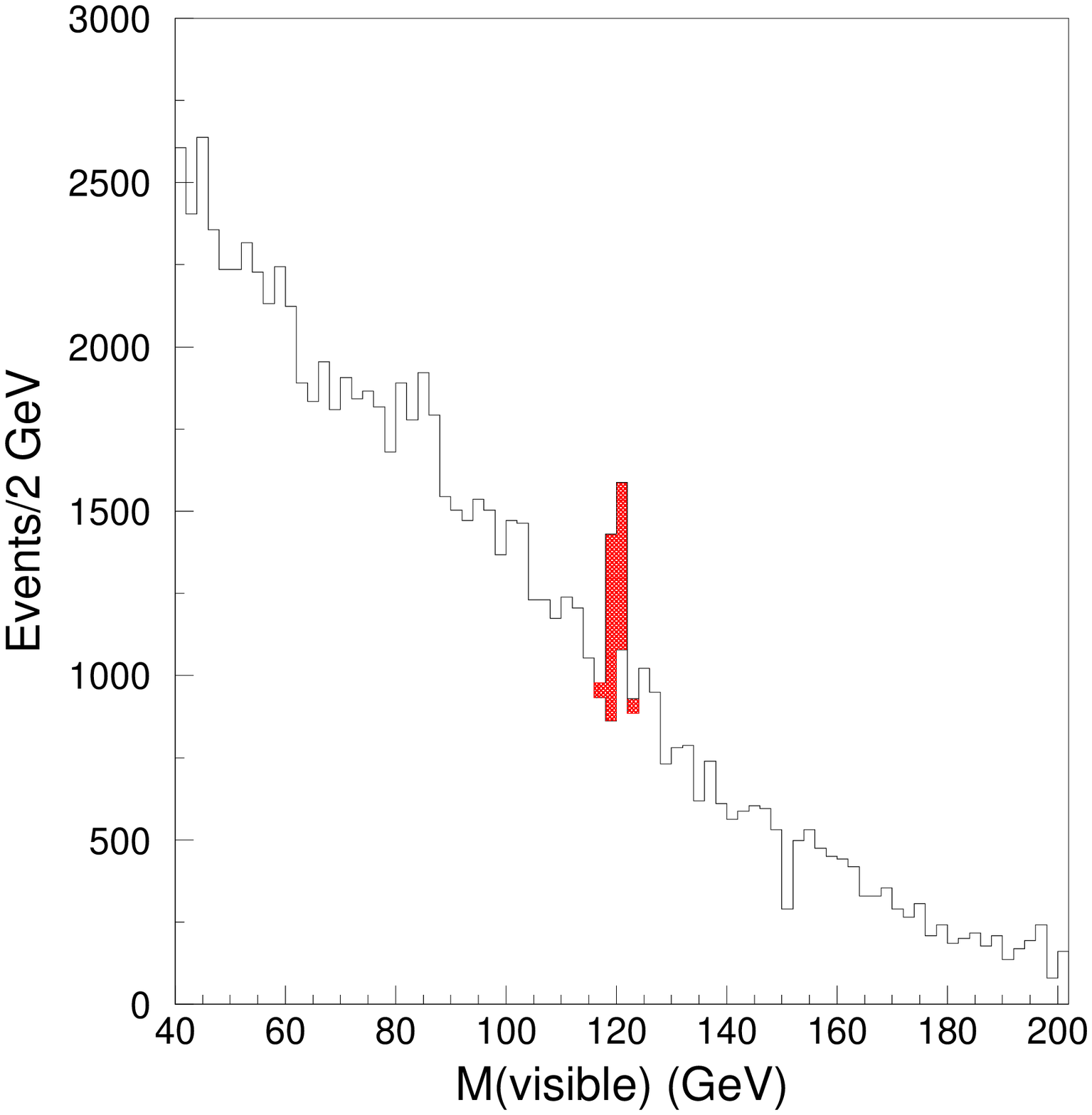 ,width=6.0cm}}
     \end{tabular}
     \end{center}
     \caption{Histograms of the visible mass following $\ee\ra\nu_e\bar{\nu}_e H$  selection cuts at $\sqrt{s}=1000$~GeV for the $b\bar{b}$ decay mode and $M_H=200$~GeV (left) and 
for the $\gamma\gamma$ decay mode and $M_H=120$~GeV (right).
The histograms contain
non-Higgs SM background (white), signal Higgs decay (red) and non-signal Higgs decays (green).
              }
     \label{fig:higgsonetev}
     \end{figure}

Consider, for example, the $b\bar{b}$ decay of a 200~GeV Higgs boson, which
is inaccessible at $\sqrt{s}=350$~GeV, and the $\gamma\gamma$ decay of a 120~GeV Higgs boson, whose branching
fraction can be measured with a relative accuracy of 25\% at $\sqrt{s}=350$~GeV with 500~fb$^{-1}$ luminosity~\cite{Boos:2000bz}.
The visible mass distributions for these two scenarios are displayed in Figure~\ref{fig:higgsonetev}
assuming 
$\sqrt{s}=1000$~GeV, 1000~fb$^{-1}$ luminosity, -80\% initial electron polarization and +50\% initial positron polarziation.
A measurement of the cross-section times branching ratio 
leads to relative branching ratio errors of
9\% for the $b\bar{b}$ decay of a 200~GeV Higgs boson and
5\% for the $\gamma\gamma$ decay of a 120~GeV Higgs boson.

\section{LHC/ILC Complementarity}

\subsection{Top Yukawa coupling}

The top Yukawa coupling $g_{ttH}$ will be probed at the LHC by measuring the cross-section for $gg\ra t\bar{t}H\ra t\bar{t}b\bar{b},\ t\bar{t}W^+W^-$.  
When ILC Higgs branching ratio measurements are combined with LHC cross-section measurements, the top Yukawa coupling
can be measured with a relative accuracy of $\Delta g_{ttH}/g_{ttH}=0.13-0.17$ for Higgs boson masses between 120 and 200~GeV~\cite{Desch:2004kf}.  

At $\sqrt{s}=800$~GeV the
top Yukawa coupling can be probed directly at the ILC using $\ee\ra t\bar{t} H$.  A relative accuracy of $\Delta g_{ttH}/g_{ttH}=0.06-0.13$ 
can be achieved for  Higgs boson masses between 120 and 200~GeV assuming  $\sqrt{s}=800$~GeV and 
1000~fb$^{-1}$ luminosity~\cite{Gay:2004}.

\subsection{Consistency test of the SUSY Higgs system and $A_t$ measurement}

\begin{figure}[ht]
     \begin{center}
     \begin{tabular}{cc}
     \mbox{\epsfig{file=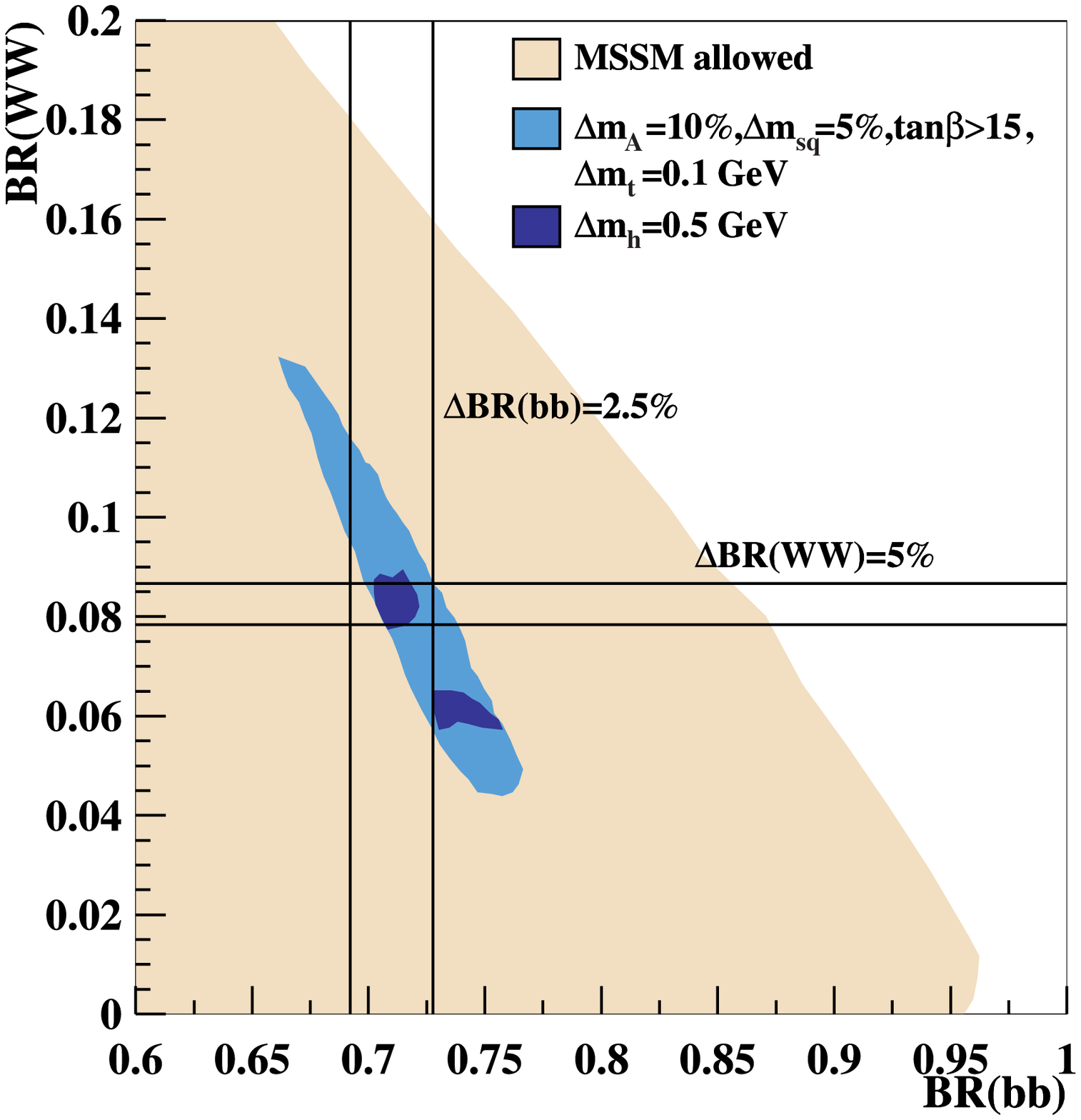,width=5.7cm}}&
     \mbox{\epsfig{file=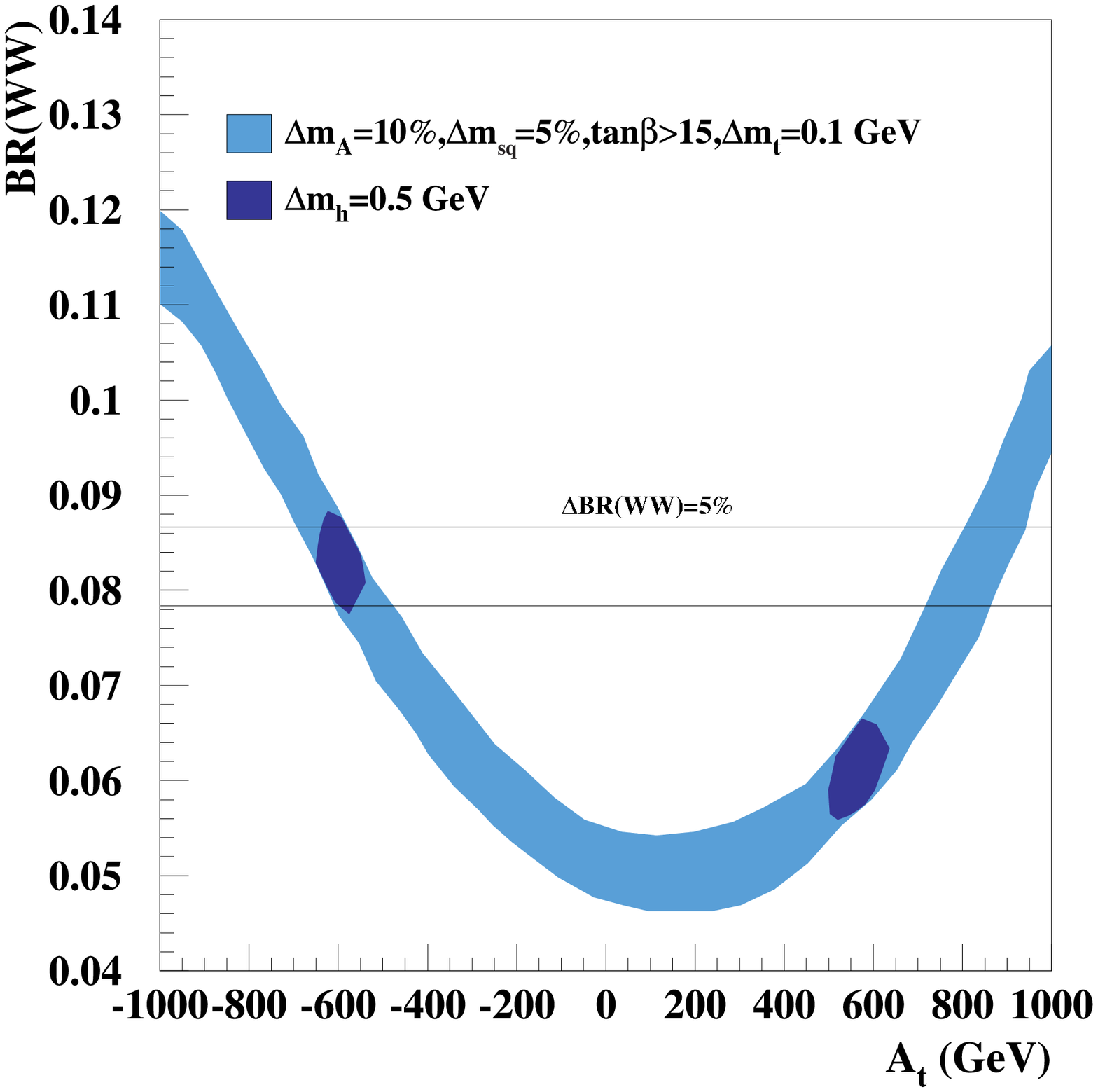 ,width=5.7cm}}
     \end{tabular}
     \end{center}
     \caption{Theoretical predictions for the allowed values of the branching ratios of the lightest MSSM Higgs boson to $WW^*$ and $b\bar{b}$ under different experimental assumptions (left),
            and predictions for the allowed values of the 
       branching ratio of the lightest MSSM Higgs boson to $WW^*$ and the trilinear coupling $A_t$ (right).  Also shown is the  expected branching ratio precision from ILC measurements. }
     \label{fig:susyhiggs}
     \end{figure}

A  recent study provides examples of how precision Higgs boson measurements at the ILC can be combined with LHC measurements of the masses
of SUSY particles to test supersymmetric relationships and extract electroweak scale SUSY parameters~\cite{Desch:2004cu}.  
In one example, LHC measurements of the masses of the pseudoscalar Higgs $A$,
the bottom squarks $\tilde{b}_1$, $\tilde{b}_2$,  and top squarks $\tilde{t}_1$, $\tilde{t}_2$ are combined with ILC measurements of the masses of the top quark
and lightest Higgs boson to predict the branching ratios of the lightest Higgs boson to $b\bar{b}$ and $WW^*$.  The dark blue splotches in Figure~\ref{fig:susyhiggs}(a) indicate
the allowed regions for the Higgs branching ratios to  $b\bar{b}$ and $WW^*$, while the bands for the ILC's Higgs branching ratio measurements show how
well these predictions will be tested.  If the ILC branching ratio measurements are consistent with the MSSM predictions then the branching ratio measurements
also provide an indirect measurement of the trilinear coupling $A_t$.

\section{Conclusion}

Table~\ref{tab:brs} summarizes the accuracy with which Higgs branching ratios, the total Higgs decay width, the top Yukawa coupling
and the Higgs self-coupling $\lambda$  can be measured at the ILC through a combination of
500~fb$^{-1}$ luminosity at $\sqrt{s}=350$~GeV and 
1000~fb$^{-1}$ luminosity at $\sqrt{s}=1000$~GeV~\cite{Yamashita:2004,Barklow:2003hz,Gay:2004,Battaglia:1999re,Brient:2002}.

\begin{table}[htbp]
\center{\begin{tabular}{l|rrrr}
   & \multicolumn{4}{c}{Higgs Mass (GeV)} \\
   & 120 & 140 & 160 & 200 \\ \hline
 $\Delta B_{bb}/B_{bb}$ &  $ 0.016$  & $ 0.018$    & $ 0.020$  & $ 0.090$   \\
 $\Delta B_{WW}/B_{WW}$ &  $ 0.020$  & $ 0.018$    & $ 0.010$  & $ 0.025$   \\
 $\Delta B_{gg}/B_{gg}$ &  $ 0.023$  & $ 0.035$    & $ 0.146$  &               \\
 $\Delta B_{\gamma\gamma}/B_{\gamma\gamma}$ &  $ 0.054$  & $ 0.062$    & $ 0.237$  &               \\
 $\Delta B_{\tau\tau}/B_{\tau\tau}$ &  $ 0.050$  & $ 0.080$    &   &               \\
 $\Delta B_{cc}/B_{cc}$ &  $ 0.083$  & $ 0.190$    &   &               \\
 $\Delta \Gamma_{tot}/\Gamma_{tot}$ &  $ 0.034$  & $ 0.036$    & $ 0.020$  & $ 0.050$   \\
 $\Delta g_{ttH}/g_{ttH}$ & $ 0.060$  & $ 0.090$    & $ 0.080$  & $ 0.130$   \\
 $\Delta \lambda/\lambda$  &  $ 0.120$  &              &   &               \\
\end{tabular}
\caption{\label{tab:brs}\it 
Relative accuracies for the measurement of Higgs branching ratios, the Higgs boson total decay width, the top Yukawa coupling $g_{ttH}$ and
the triple Higgs coupling $\lambda$ obtained through a combination of 500~fb$^{-1}$ luminosity at $\sqrt{s}=350$~GeV and 
1000~fb$^{-1}$ luminosity at $\sqrt{s}=1000$~GeV.
}}
\end{table}

In summary, 
there has been progress in understanding beam related systematic errors in the measurement of Higgs
boson masses at the ILC.  
It has also been shown that Higgs physics research at $\sqrt{s}=1000$~GeV produces improvements in Higgs branching ratio and self-coupling 
measurements, 
even if the Higgs boson mass is less than 200~GeV.  Finally, it is not hard
to find examples of LHC/ILC complementarity in studies of Higgs and SUSY physics.  The ILC's few hundred MeV Higgs mass measurement and
few percent branching ratio measurements are used repeatedly in these studies.


\end{document}